\documentclass[journal]{IEEEtran}

\usepackage[T1]{fontenc}
\usepackage{graphicx}
\usepackage{amsmath,amssymb,bm}
\usepackage{algorithm}
\usepackage{algpseudocode}
\usepackage{xcolor}
\usepackage{url}
\usepackage[hypertexnames=false]{hyperref}
\hypersetup{colorlinks=true, linkcolor=blue, citecolor=blue, urlcolor=blue}

\newtheorem{theorem}{Theorem}
\newtheorem{proposition}{Proposition}
\newtheorem{remark}{Remark}

\begin{document}

\title{On Multi-Constrained Path Finding utilizing Grover Search and Decoded Quantum Interferometry}

\author{Dmitrii~Chemodanov%
\thanks{D. Chemodanov, e-mail: duman190 AT gmail.com.}}

\maketitle

\begin{abstract}
Finding a path whose totals under several additive metrics all fall inside prescribed ranges is weakly NP-complete. We ask what two quantum algorithms can do for it, against strong classical baselines. For directed acyclic graphs (DAGs) we give a Grover search whose state preparation builds the uniform superposition over all source-to-destination paths with one qubit per arc and no ancilla. We prove it correct, give its exact gate count, and show that measured strings form an error-detecting code whose distance is a girth-like graph parameter. We complete the oracle and count its gates: one search on our largest instance, 136 arc qubits and 262{,}144 paths, needs about $4.4\times10^{5}$ Toffoli gates, while classical dynamic programming decides the same instance with 18{,}754 labels. Grover search needs $O(\sqrt{N/M})$ oracle calls for $M$ good paths among $N$. Classical meet-in-the-middle matches this exponent in the worst case, and known quantum algorithms with quantum memory beat it. For a staged fabric with node-independent linear weights, maximizing the number of satisfied metrics is max-LINSAT, and with power-of-index weights it is Optimal Polynomial Intersection, where no known polynomial-time classical algorithm matches decoded quantum interferometry (DQI). We verify this translation in simulation and run two small binary instances on IBM Heron processors, which keep 67\,\% and 15\,\% of the ideal gain over uniform sampling.
\end{abstract}


\section{Introduction}

Many network services need a path that meets several requirements at once: a delay budget, a loss budget, a hop limit, a cost limit. With two or more additive upper-bound constraints the question of whether such a path exists is NP-complete~\cite{wang1996,garey1979}. With two-sided ranges one metric suffices: on a chain of stages offering an arc of weight $a_{j}$ or $0$, a path of total exactly $T$ is a subset-sum solution. Both versions are weakly NP-complete for a fixed number $K$ of metrics, since dynamic programming over nodes and partial totals runs in time pseudo-polynomial in the weights~\cite{jaffe1984,hassin1992}. The hard instances therefore have large weights or many metrics. Practical systems use approximation schemes~\cite{hassin1992}, Lagrangian relaxation~\cite{juttner2001}, heuristics~\cite{korkmaz2001} and exact search with pruning~\cite{vanmieghem2004,kuipers2002,chemodanov2019}.

Quantum computers are often proposed for problems of this kind. This paper works out two concrete routes, states what each one delivers, and sets each against the classical methods that matter for it.

The first route assumes only acyclicity and integer weights. It is amplitude amplification~\cite{brassard2002} over the set of paths, with a state preparation that never leaves that set. Plain Grover search needs $O(\sqrt{N/M})$ oracle calls for $M$ paths in range among $N$. Whether this helps depends on $M$, on the structure of the graph, and on the classical method it is compared with. Section~\ref{sec:cost} makes the comparison precise: in the worst case over instances classical meet-in-the-middle reaches the same exponent, known quantum algorithms do better with quantum random-access memory, and on the instances we test, which are in the pseudo-polynomial regime, dynamic programming is far cheaper than any of them.

The second route gives up generality. Decoded quantum interferometry (DQI)~\cite{jordan2025} turns an optimization problem into a decoding problem by a quantum Fourier transform. For one algebraically structured problem, Optimal Polynomial Intersection (OPI), it reaches approximation ratios that no known polynomial-time classical algorithm reaches. We show which staged fabrics turn the many-metric path problem into max-LINSAT, the problem DQI addresses, and that OPI is the case of power-of-index weights. The price is that weights must not depend on the node, totals wrap around modulo $p$, and the goal becomes satisfying as many metrics as possible.

\textbf{Contributions.}
(i) An ancilla-free state preparation $U_{S}$ for the uniform superposition over the $s$--$t$ paths of a DAG, with a proof, an exact CNOT count, depth, connectivity and precision requirements, and the distance of the flow-conservation code it induces (Theorem~\ref{thm:prep}, Proposition~\ref{prop:distance}).
(ii) A complete range-constrained oracle, including uncomputation, with a logical Toffoli and qubit estimate per oracle call and per search (Table~\ref{tab:resources}).
(iii) A comparison of classical and quantum methods as functions of $N$, $M$, depth, degree and weights (Table~\ref{tab:complexity}), evaluated on 22 instances against exact dynamic programming, exact and sub-sampled meet-in-the-middle, uniform sampling and a classical Boltzmann sampler.
(iv) A characterization of the staged fabrics on which the path problem is max-LINSAT, with OPI as the power-of-index case; a state-level simulation of DQI on six OPI instances; and runs of two binary instances on IBM hardware, re-analyzed with both weight choices and a noise model on the transpiled circuits.
(v) A statement of what is not shown (Section~\ref{sec:limits}).


\section{Related Work}

\subsection{Classical Multi-Constrained Path Finding}

Wang and Crowcroft showed that two additive upper-bound constraints make path selection NP-complete~\cite{wang1996}, and Jaffe gave pseudo-polynomial and approximation algorithms~\cite{jaffe1984}. The restricted shortest path problem has a fully polynomial approximation scheme~\cite{hassin1992}. Lagrangian relaxation (LARAC) gives fast heuristics with a bound on the gap~\cite{juttner2001}, and H\_MCOP adds look-ahead~\cite{korkmaz2001}. Exact algorithms such as SAMCRA use dominance pruning and look-ahead~\cite{vanmieghem2004}; Kuipers et al.\ survey the field~\cite{kuipers2002}. In operations research the same problem is the resource-constrained shortest path, solved by bidirectional labeling that meets in the middle~\cite{righini2006}, an idea that goes back to Horowitz and Sahni~\cite{horowitz1974}.

Our earlier work treats the constrained shortest path as the building block of virtual network services~\cite{chemodanov2016,chemodanov2019}. Its neighborhoods method builds hop-bounded neighborhoods from the source and from the destination and joins them, and reports a quadratically lower upper bound and an order of magnitude less running time than exhaustive breadth-first search. That bound is an informal counting argument. The rigorous version for path constraints is meet-in-the-middle over partial-path totals with a range-query join, which we state in Section~\ref{sec:cost}. Requests with several QoS and QoE targets arise in the same line of work~\cite{chemodanov2021,chemodanov2019infocom,chemodanov2020}.

\subsection{Quantum Search, State Preparation and Path Problems}

Grover search finds one of $M$ marked items among $N$ with about $(\pi/4)\sqrt{N/M}$ oracle calls~\cite{grover1996,boyer1998}, which is optimal for unstructured search~\cite{bennett1997}. The number of rounds can be chosen without knowing $M$~\cite{boyer1998}, the uniform start state can be replaced by any state a circuit prepares~\cite{brassard2002}, and a minimum can be found by repeated search with a falling threshold~\cite{durr1996}. Grover mixers use the reflection about a prepared state inside QAOA~\cite{baertschi2020}, and some mixers preserve flow conservation~\cite{zhang2021}. Jeffery, Kimmel and Piedrafita sample edges on $s$--$t$ paths with a span-program state~\cite{jeffery2023}.

The state preparation of Section~\ref{sec:grover} is a relative of three known constructions: Grover--Rudolph preparation of efficiently integrable distributions by conditional rotations~\cite{grover2002}, here with path counts as the integrals; cascades for W and Dicke states~\cite{baertschi2019dicke}, here one cascade per node; and uniform-state preparation by Boolean methods and from decision diagrams~\cite{mozafari2021,mozafari2022}, here with the DAG itself as the diagram. What we add is the ancilla-free two-local form on the line graph of the DAG, its exact cost, and the code it induces.

Quantum algorithms with better exponents than Grover search exist for problems this one contains. Quantum walks find claws and collisions in $O(N^{1/3})$ queries~\cite{ambainis2007,tani2009}; quantum subset-sum algorithms beat the best classical exponents~\cite{bonnetain2020}; and quantum backtracking speeds up tree search with pruning quadratically in the tree size~\cite{montanaro2018}. Babbush et al.\ argue that quadratic speedups are unlikely to survive error correction on early fault-tolerant machines~\cite{babbush2021}, and Sanders et al.\ compile Grover-type optimization oracles to fault-tolerant cost~\cite{sanders2020}.

\subsection{Decoded Quantum Interferometry and Where It Applies}

DQI~\cite{jordan2025} addresses max-LINSAT: given linear functions over a finite field and an allowed set for each, find the input that puts the most functions inside their sets. When the functions are evaluations of a polynomial, the problem is OPI and the decoder needed is a Reed--Solomon decoder. On unstructured sparse instances DQI faces a spin-glass obstruction~\cite{anschuetz2025}, so its promise lies with algebraic structure.

\emph{Verifiable advantage.} A proposed OPI solution is checked by counting satisfied constraints. Khattar et al.\ give optimized circuits and estimate about $5.72$ million Toffoli gates and 1{,}885 logical qubits for an instance over $\mathrm{GF}(2^{12})$ ($m=4095$, $n=70$) that they estimate needs more than $10^{23}$ classical trials~\cite{khattar2025}. A separation has been proven for a folded variant of OPI in an oracle model~\cite{kramer2026}.

\emph{Limits and extensions.} Contiguous allowed sets give noisy polynomial reconstruction, which lattice methods break in one parameter regime~\cite{bleichenbacher2000}. DQI is not optimal on worst-case OPI~\cite{sun2026,horinaga2026}, and it has been extended to algebraic-geometry codes~\cite{gu2025}.

We know of no routing application with power-of-index modular metrics. Section~\ref{sec:dqi} says which weight structures DQI needs and names periodic scheduling as a candidate.

\section{Problem Statement}

\subsection{General Case: Range-Constrained Paths on a DAG}

Let $G=(V,E)$ be a directed acyclic graph with source $s$ and destination $t$, pruned so that every node lies on an $s$--$t$ path, and without parallel arcs unless stated. Each arc $e$ carries $K$ non-negative integer weights $w_{1}(e),\dots,w_{K}(e)$. The total of metric $i$ along a path $P$ is
\begin{equation}
  T_{i}(P) = \sum_{e \in P} w_{i}(e),
  \label{eq:total}
\end{equation}
and each metric has a range $[L_{i},U_{i}]$. The task is to find an $s$--$t$ path with
\begin{equation}
  L_{i} \le T_{i}(P) \le U_{i} \qquad \text{for every } i = 1,\dots,K,
  \label{eq:ranges}
\end{equation}
or to report that none exists. A lower bound of zero gives the usual upper-bound form, and minimizing one metric under bounds on the others reduces to a sequence of such feasibility questions. We write $N$ for the number of $s$--$t$ paths and $M$ for the number in range, $d$ for the largest number of arcs on a path, $\Delta$ for the largest in- or out-degree, and $W_{i}$ for the largest path total of metric $i$. The count $N(v)$ of $v$--$t$ paths is computed for all $v$ in $O(|E|)$ time, and $N = N(s) \le \Delta^{d}$.

The average out-degree $|E|/|V|$ does not bound $N$: with $s$ joined to $k$ middle nodes and each joined to $t$, $N = k$ while $(|E|/|V|)^{d} < 4$.

\subsection{Special Case: A Staged Linear Fabric}
\label{sec:fabric}

\emph{Fabric.} $n$ stages in series between source and destination. Every stage offers $q$ parallel ports labeled by the elements of the finite field $\mathbb{F}_{q}$. A path is one port per stage, $x=(x_{0},\dots,x_{n-1})\in\mathbb{F}_{q}^{n}$, so $N=q^{n}$.

\emph{Metrics.} There are $m$ metrics. Under metric $y$, port $a$ at stage $j$ weighs $a\,B[y,j]$ for a fixed matrix $B\in\mathbb{F}_{q}^{m\times n}$, so the totals are
\begin{equation}
  T_{y}(x) = \sum_{j=0}^{n-1} B[y,j]\,x_{j} = (Bx)_{y}.
  \label{eq:modtotal}
\end{equation}

\emph{Constraints and goal.} Each metric has a set $F_{y}$ of $r$ allowed totals. Find the path that satisfies the most metrics. This is max-LINSAT.

Two members matter here. \emph{OPI}: $q=p$ prime, $m=p-1$ metrics $y=1,\dots,p-1$, and $B[y,j]=y^{j}$, so a port's weight is $a\,y^{j}\bmod p$ and the total of metric $y$ is a polynomial evaluated at $y$. \emph{Binary fabrics}: $q=2$, a total is a parity and the problem is max-XORSAT; the hardware instances of Section~\ref{sec:hw} are of this kind.

The special case differs from the general one in three ways. Every combination of ports is a path, so the graph is a chain of bundles of parallel arcs. Totals are taken in $\mathbb{F}_{q}$. And the goal is the largest number of satisfied metrics, since with $m\gg n$ a path that satisfies all of them usually does not exist.

\begin{figure}[t]
  \centering
  \includegraphics[width=\columnwidth]{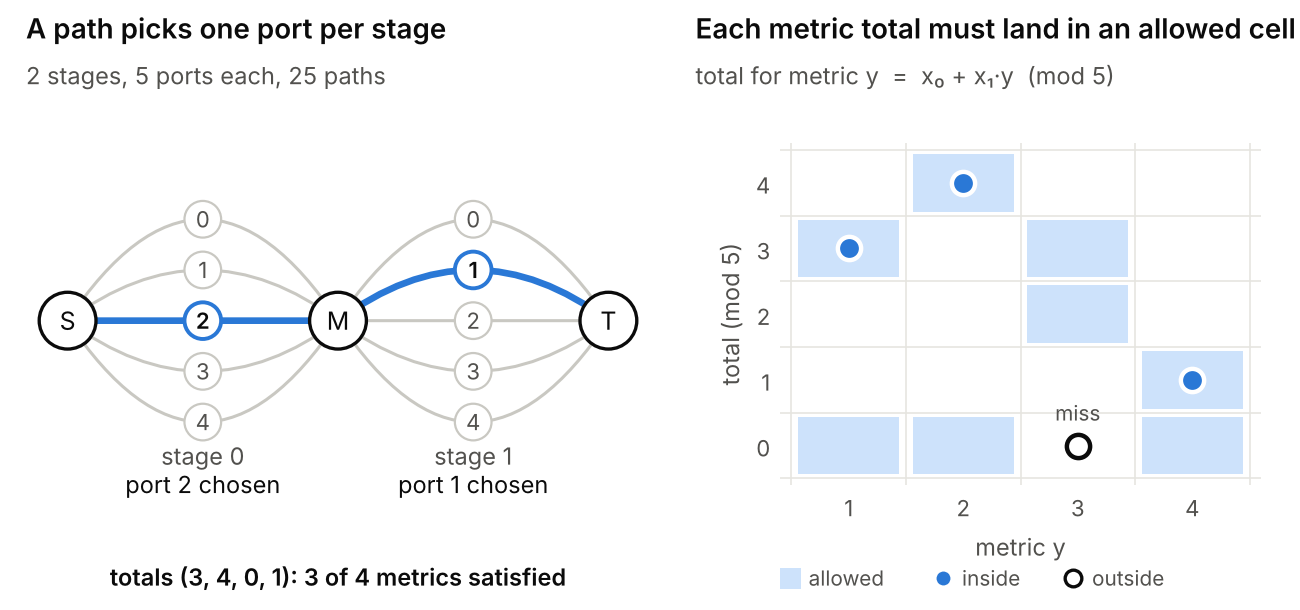}
  \caption{The smallest OPI instance: two stages, five ports per stage, four metrics, two allowed totals per metric. The highlighted path takes port 2, then port 1. Its totals are 3, 4, 0 and 1, and three of them are allowed. No path satisfies all four.}
  \label{fig:fabric}
\end{figure}

\section{Grover Search on General DAGs}
\label{sec:grover}

\subsection{State Preparation}

Each arc is one qubit, and a path is the bit string with ones on its arcs. Number the out-arcs of each node $v\neq t$ as $e_{1},\dots,e_{o}$ and write $h(e)$ for the head of arc $e$.

\emph{Construction of $U_{S}$.} Visit the nodes $v\neq t$ in topological order. (a) \emph{Token.} If $v=s$, apply $X$ to $e_{1}$. Otherwise apply a CNOT from every in-arc of $v$ onto $e_{1}$. (b) \emph{Split.} For $j=1,\dots,o-1$, apply a controlled $R_{y}(\theta_{j})$ with control $e_{j}$ and target $e_{j+1}$, then a CNOT from $e_{j+1}$ to $e_{j}$, where
\begin{equation}
  \cos^{2}\frac{\theta_{j}}{2} = \frac{N(h(e_{j}))}{\sum_{l\ge j} N(h(e_{l}))}.
  \label{eq:angles}
\end{equation}

\begin{theorem}\label{thm:prep}
$U_{S}$ maps the all-zero string to
\begin{equation}
  U_{S}\lvert 0\rangle = \frac{1}{\sqrt{N}}\sum_{P} \lvert P\rangle ,
  \label{eq:prep}
\end{equation}
the sum running over all $s$--$t$ paths. It uses no ancilla qubit and $4|E| - \operatorname{indeg}(t) - 3|V| + 3$ CNOT-equivalents, counting a controlled rotation as two CNOTs.
\end{theorem}

\begin{IEEEproof}
Call a \emph{prefix} an $s$-path whose arcs all leave processed nodes and whose last arc enters an unprocessed node. We show by induction over the topological order that after processing a set of nodes the state is $\sum_{\pi}\sqrt{N(h(\pi))/N}\,\lvert\pi\rangle$, where $\pi$ runs over the prefixes and $h(\pi)$ is the head of the last arc. Before any node is processed the only prefix is the empty path at $s$, with $N(s)/N=1$. When $v$ is processed, a branch whose prefix ends at $v$ has exactly one in-arc of $v$ set, so step (a) sets $e_{1}$; XOR equals OR here. Step (b) is a W-type cascade: by (\ref{eq:angles}) it moves the set bit to $e_{j}$ with amplitude $\sqrt{N(h(e_{j}))/N(v)}$, and the product with $\sqrt{N(v)/N}$ is $\sqrt{N(h(e_{j}))/N}$. A branch whose prefix ends elsewhere has no in-arc of $v$ set, because such an arc would end the prefix at $v$, and $e_{1},\dots,e_{o}$ are still zero, so (a) and (b) act trivially. When all nodes but $t$ are processed every prefix is a complete path and $N(t)=1$. A node with $i$ in-arcs and $o$ out-arcs costs $i + 3(o-1)$ CNOT-equivalents; summing, $\sum_{v\neq s,t}\operatorname{indeg}(v) + 3\sum_{v\neq t}(\operatorname{outdeg}(v)-1) = (|E|-\operatorname{indeg}(t)) + 3(|E|-|V|+1)$.
\end{IEEEproof}

\emph{Depth, connectivity and precision.} Every gate acts on two arcs that share a node, so the circuit is two-local on the line graph of $G$. Nodes with no arc between them act on disjoint qubits, so nodes in the same layer of a longest-path layering run in parallel and the depth is $O(d\,\Delta)$ two-qubit layers. The circuit has $R = |E|-|V|+1$ controlled rotations. Errors in operator norm add, so synthesizing each rotation to accuracy $\varepsilon = \delta/(2R)$ keeps the output within total variation distance about $\delta$; in Clifford+T that costs about $3\log_{2}(1/\varepsilon)$ T gates per rotation~\cite{ross2016}. For I5 ($R=39$) and for the 136-arc layered instance ($R=99$), $\delta=10^{-3}$ needs $\varepsilon\approx 1.3\times10^{-5}$ and $5\times10^{-6}$, that is about 16 and 18 bits of angle. This covers one application of $U_{S}$. A search with $k$ rounds applies $U_{S}$ or its inverse $2k+1$ times, so $\varepsilon$ shrinks by that factor: 185 applications on the 136-arc instance, about 8 more bits. The angles themselves are ratios of path counts and are computed classically with $O(\log N)$-bit integers.

Measuring $U_{S}\lvert 0\rangle$ is a classical random walk from $s$ that branches in proportion to the number of onward paths, so state preparation alone has no quantum content.

\begin{proposition}\label{prop:distance}
Let $g(G)$ be the smallest total length of two distinct internally disjoint directed $u$--$v$ routes, over all node pairs $u,v$. The Hamming distance between the strings of two distinct $s$--$t$ paths is at least $g(G)$, with equality for some pair. Without parallel arcs $g(G)\ge 3$.
\end{proposition}

\begin{IEEEproof}
Two distinct paths split at some node $u$ and next meet at some $v$; the two segments between them are internally disjoint and contribute their total length to the distance, and further detours only add. Conversely, two such routes for the minimizing pair, extended by a common $s$--$u$ prefix and $v$--$t$ suffix, which exist after pruning, give two paths at distance $g(G)$. Without parallel arcs at least one of the two routes has two arcs.
\end{IEEEproof}

Any one or two bit flips of a path string therefore give a string that is not a path; the strings with valid flow conservation form a code of distance $g(G)$. Subdividing arcs raises $g(G)$ at the cost of qubits. The fabric of the special case has parallel arcs and $g=2$.

\subsection{Oracle}
\label{sec:oracle}

The oracle marks the paths in range. It uses one total register of $b_{i}=\lceil\log_{2}(W_{i}+1)\rceil$ bits per metric, $2K$ test qubits and a flag in the state $\lvert -\rangle$.

\begin{algorithm}[t]
\caption{Range-constrained Grover oracle}\label{alg:oracle}
\begin{algorithmic}[1]
\Require Arc qubits $\lvert e\rangle$; registers $\lvert T_{i}\rangle$ and test qubits $\lvert c_{i}^{\pm}\rangle$ set to zero; flag $\lvert -\rangle$
\Ensure Phase $-1$ on every path with $L_{i} \le T_{i} \le U_{i}$ for all $i$; all work registers returned to zero
\For{every arc $e$ and $i = 1$ to $K$}
    \State $\lvert T_{i}\rangle \leftarrow \lvert T_{i}+ w_{i}(e)\rangle$ controlled on $\lvert e\rangle$
\EndFor
\For{$i = 1$ to $K$}
  \State $c_{i}^{-} \leftarrow [T_{i} \ge L_{i}]$, \ $c_{i}^{+} \leftarrow [T_{i} \le U_{i}]$
\EndFor
\State flip the flag if all $2K$ test qubits are 1
\For{$i = K$ down to $1$}
  \State uncompute $c_{i}^{+}$ and $c_{i}^{-}$
\EndFor
\For{every arc $e$ and $i = K$ down to $1$, in reverse order}
    \State $\lvert T_{i}\rangle \leftarrow \lvert T_{i}- w_{i}(e)\rangle$ controlled on $\lvert e\rangle$
\EndFor
\end{algorithmic}
\end{algorithm}

\emph{Resource count.} We count logical gates with the following constructions. A controlled addition of a constant loads the constant into a scratch register with CNOTs controlled on the arc qubit and adds it with Gidney's adder, $b_{i}-1$ Toffoli gates with measurement-based uncomputation of its logical ANDs~\cite{gidney2018}; subtraction costs the same. A comparison with a constant costs $b_{i}$ Toffoli gates to compute, and we assume its uncomputation is measurement-based like the adder's and costs none; with coherent uncomputation the comparator term of (\ref{eq:toffoli}) doubles, which adds $2\sum_{i}b_{i}$ Toffoli gates per call, 24 to the 119 of I1 and 40 to the 4{,}669 of the largest instance; the $2K$-input AND costs $2K-1$; the reflection about the all-zero string on $|E|$ qubits costs $|E|-1$. One oracle call then costs
\begin{equation}
  2|E|\sum_{i}(b_{i}-1) + 2\sum_{i} b_{i} + 2K - 1
  \label{eq:toffoli}
\end{equation}
Toffoli gates, and one round adds $|E|-1$ for the reflection and $4R$ arbitrary rotations for $U_{S}^{\dagger}$ and $U_{S}$. The logical qubits are $|E|+\sum_{i}b_{i}+1$ for arcs, totals and flag, plus $\max_{i}b_{i}$ scratch, $\max_{i}b_{i}-1$ adder ancillas, $2K$ test qubits and $\max(2K-2,|E|-2)$ ancillas for the wide ANDs, reused across steps. Table~\ref{tab:resources} evaluates these counts. No error-correction overhead is included.

\subsection{Diffusion and Number of Rounds}

The diffusion step reflects about the prepared state: apply $U_{S}^{\dagger}$, flip the sign of the all-zero string, apply $U_{S}$. Because the reflection is about the superposition of paths, the ideal state never leaves their span. After $k$ rounds a measurement returns a path in range with probability~\cite{boyer1998,brassard2002}
\begin{equation}
  P(k) = \sin^{2}\bigl((2k+1)\theta\bigr), \qquad
  \sin\theta = \sqrt{\frac{M}{N}},
  \label{eq:sinelaw}
\end{equation}
and the standard choice of rounds is
\begin{equation}
  k^{*} = \left\lfloor \frac{\pi}{4\theta} \right\rfloor
  \approx \frac{\pi}{4}\sqrt{\frac{N}{M}} .
  \label{eq:rounds}
\end{equation}
It need not maximize $P(k)$: on I2 one round gives 0.784 and three give 0.994. When $M$ is unknown, rounds are drawn from a growing interval and the expected number of oracle calls stays $O(\sqrt{N/M})$; certifying that no path is in range takes $\Theta(\sqrt{N})$ calls with one-sided error~\cite{boyer1998}. To minimize a metric, its upper limit is lowered after every success~\cite{durr1996}.

\begin{figure*}[t]
  \centering
  \includegraphics[width=0.82\textwidth]{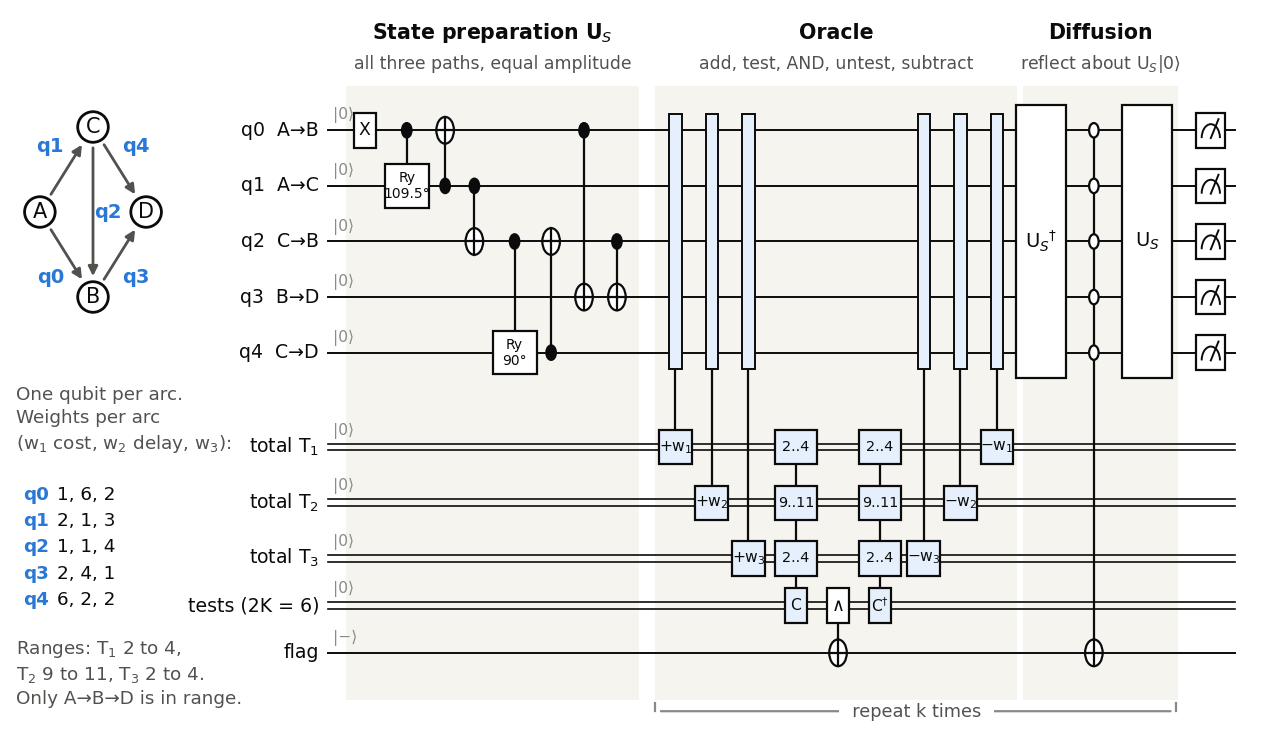}
  \caption{One Grover round on instance I1 (four nodes, five arcs, three paths, three metrics). The eight gates of $U_{S}$ are drawn exactly. In the oracle, a tall bracket on the arc wires joined to a box $+w_{i}$ stands for five singly-controlled additions, one per arc qubit $q_{e}$, of $w_{i}(e)$ into total $T_{i}$; it is not a five-fold controlled gate. The comparators $C$ write six test bits, their conjunction $\wedge$ flips the flag, and $C^{\dagger}$ and the subtractions return all work registers to zero. Open circles are controls on 0; the reflection about $\lvert 0\rangle$ on the five arc qubits is drawn with the flag as phase target. The drawing shows the 18 qubits of arcs, totals and flag and the six test qubits; adder scratch and AND ancillas are counted in Table~\ref{tab:resources} but not drawn.}
  \label{fig:circuit}
\end{figure*}

\subsection{Cost Against Classical and Quantum Alternatives}
\label{sec:cost}

Table~\ref{tab:complexity} collects the methods. Four points summarize it.

\emph{Worst case.} Classical meet-in-the-middle stores the totals of all forward partial paths of $\lceil d/2\rceil$ arcs in a $K$-dimensional range tree and, for each backward partial path, asks whether some forward one completes the totals into the ranges. This takes $\tilde O(\Delta^{\lceil d/2\rceil})$ time and memory for fixed $K$~\cite{horowitz1974,righini2006}. On layered DAGs that is $\tilde O(\sqrt N)$, the same exponent as Grover search with $M=1$. The one asymptotic difference is memory: Grover search needs $O(|E| + \sum_{i}b_{i})$ qubits.

\emph{Many solutions.} In the worst case the better of uniform sampling and meet-in-the-middle costs $\tilde O(\min(N/M,\sqrt N))$ and Grover search $O(\sqrt{N/M})$, so Grover search can be ahead by up to $N^{1/4}$, at $M\approx\sqrt N$. Structure can remove that gap. On layered DAGs, any forward half joined to any backward half is a path, and sampling $s$ halves on each side tests $s^{2}$ paths; with a range-query join this costs $\tilde O(s)$ time, and $s\approx\sqrt{N/M}$ suffices when solutions are spread out. Fig.~\ref{fig:groverresults} measures this: 560 half-path evaluations reach 90\,\% success on the largest layered instance, a log-log slope of 0.49 against $N/M$.

\emph{Better quantum exponents.} Tabulating prefixes of about $d/3$ arcs classically and running Grover search over suffixes, with a range query into the table as the oracle, costs $\tilde O(\Delta^{d/3})$ time with $\Delta^{d/3}$ entries of quantum random-access memory; $U_{S}$ restricted to the suffix sub-DAGs is the state preparation it needs. For exact targets ($L_{i}=U_{i}$) quantum-walk claw finding needs $O(N^{1/3})$ queries~\cite{tani2009,ambainis2007}, and ranges reduce to $\prod_{i}(U_{i}-L_{i}+1)$ exact targets, 27 on our instances. For subset sum, which the problem contains, the best quantum exponent is below the best classical one~\cite{bonnetain2020}. Quantum backtracking over the pruned tree of partial paths~\cite{montanaro2018} is the quantum counterpart of branch-and-bound and SAMCRA. So plain Grover search over whole paths is not the best quantum approach.

\emph{Practice.} All our instances are in the pseudo-polynomial regime. Exact labeling with upper-bound pruning decides the largest one with 18{,}754 labels and counts $M$ exactly, while one Grover search needs $4.4\times10^{5}$ logical Toffoli gates and $3.6\times10^{4}$ arbitrary rotations (Table~\ref{tab:resources}), before error correction, which is expected to absorb a quadratic speedup at realistic sizes~\cite{babbush2021,sanders2020}. Arc-indicator encoding adds to the cost: a path uses at most $d$ arcs but every oracle call touches all $|E|$, and the 136 arc qubits of the largest instance could be replaced by $d\lceil\log_{2}\Delta\rceil = 20$ successor-choice qubits at the price of a weight lookup in each step.

Our conclusion is narrower than ``no quantum benefit'': the plain Grover circuit presented here ties classical meet-in-the-middle in exponent when solutions are few, can be ahead by up to $N^{1/4}$ only when they are many and not spread out, is beaten by quantum algorithms that assume quantum memory, and is far more expensive than dynamic programming on every instance we can construct and simulate.

\begin{table*}[t]
\caption{Cost of methods for range-constrained $s$--$t$ paths with $K$ fixed. Evaluations count oracle calls or partial-path evaluations; $\tilde O$ hides factors polynomial in $d$, $K$ and $\log N$.}
\label{tab:complexity}
\centering
\footnotesize
\setlength{\tabcolsep}{4pt}
\begin{tabular}{llll}
\hline \hline
\textbf{Method} & \textbf{Time / evaluations} & \textbf{Memory} & \textbf{Applies when} \\
\hline
Uniform path sampling & $O(N/M)$ & $O(|E|)$ & any DAG; path counts by DP in $O(|E|)$ \\
Label-setting DP & $O(|E|\prod_{i}(U_{i}+1))$ & same & pseudo-polynomial weights \\
Exact meet-in-the-middle, range-tree join & $\tilde O(\Delta^{\lceil d/2\rceil})$ & $\tilde O(\Delta^{\lceil d/2\rceil})$ & any DAG, split at hop $d/2$ \\
Sub-sampled meet-in-the-middle & $\tilde O(\sqrt{N/M})$ (measured) & $\tilde O(\sqrt{N/M})$ & layered DAGs, spread-out solutions \\
Grover / BBHT with $U_{S}$ & $O(\sqrt{N/M})$; $\Theta(\sqrt N)$ to certify none & $O(|E|+\sum_{i}b_{i})$ qubits & each call $\approx 2|E|\sum_{i}b_{i}$ Toffoli \\
Table of prefixes + Grover over suffixes & $\tilde O(\Delta^{d/3})$ & $\tilde O(\Delta^{d/3})$ QRAM & quantum random-access memory \\
Quantum-walk claw finding & $O(N^{1/3})$ per exact target vector & $O(N^{1/3})$ & ranges as $\prod_{i}(U_{i}-L_{i}+1)$ targets \\
Quantum backtracking~\cite{montanaro2018} & $\tilde O(\sqrt{T}\,d)$, $T$ = pruned tree size & $\mathrm{poly}(d)$ qubits & effective pruning predicate \\
\hline \hline
\end{tabular}
\end{table*}

\section{DQI on Staged Linear Fabrics}
\label{sec:dqi}

\subsection{Which Fabrics Admit DQI}

DQI needs the set of paths to be a group and every metric total to be a linear function on it. On the fabric of Section~\ref{sec:fabric} both hold: the paths are $\mathbb{F}_{q}^{n}$ and the totals are $Bx$. The following remark shows why node-dependent weights, as in any real network, break this.

\begin{remark}\label{rem:linear}
Take a layered DAG whose node at layer $j$ depends on the earlier choices, with ports labeled by $\mathbb{F}_{q}$, and let the weight of port $a$ at a node $u$ of layer $j$ be $\varphi_{u}(a)$. If two nodes $u\neq u'$ of layer $j$, reached by different choices at layer $j-1$, have increments $\varphi_{u}(a)-\varphi_{u}(0)\neq\varphi_{u'}(a)-\varphi_{u'}(0)$ for some $a$, then the change of the total when $x_{j}$ goes from $0$ to $a$ depends on $x_{j-1}$, so the total is not an affine function of $x$. Linearity therefore requires increments $a\,c_{j}$ that do not depend on the node, and then the DAG is equivalent, for this purpose, to a chain of bundles.
\end{remark}

In the special case the graph thus plays no role beyond fixing $n$ and $q$; the hardness comes from the number of metrics $m\gg n$. A natural setting where totals are linear over a cyclic group is periodic scheduling: in the periodic event scheduling problem~\cite{serafini1989} constraints are intervals modulo a period on differences of node potentials, so $B$ is an incidence matrix, the code $\ker(B^{\mathsf T})$ is the cycle space of the event graph, and its minimum distance is the girth. We do not pursue this here.

\subsection{OPI as a Path Problem}

Read an OPI path as the coefficients of a polynomial
\begin{equation}
  Q(y) = x_{0} + x_{1}y + \cdots + x_{n-1}y^{n-1}
  \label{eq:poly}
\end{equation}
over $\mathbb{F}_{p}$. The total of metric $y$ is $Q(y)$, and the problem is to find a polynomial of degree below $n$ that passes through as many allowed values as possible, which is OPI as stated in~\cite{jordan2025}. Table~\ref{tab:dictionary} gives the correspondence. It is a renaming: the port weights are the Vandermonde matrix of OPI.

\begin{table}[t]
\caption{Dictionary between the path problem and DQI}
\label{tab:dictionary}
\centering
\footnotesize
\begin{tabular}{p{0.42\columnwidth}p{0.48\columnwidth}}
\hline \hline
\textbf{Path language} & \textbf{DQI} \\
\hline
Stage $j$ and its chosen port $x_{j}$ & Coordinate $x_{j}\in\mathbb{F}_{q}$ \\
Path $x$ & Vector $x\in\mathbb{F}_{q}^{n}$ \\
Metric $y$ & Constraint $y$, a row of $B$ ($B[y,j]=y^{j}$ for OPI) \\
Total of metric $y$ & $(Bx)_{y}$, equal to $Q(y)$ for OPI \\
Allowed totals of metric $y$ & The set $F_{y}$ \\
Number of metrics satisfied & The max-LINSAT objective \\
Pattern $e$ of a few metrics with nonzero values & Error pattern for syndrome decoding of $\ker(B^{\mathsf T})$, a generalized Reed--Solomon code for OPI \\
\hline \hline
\end{tabular}
\end{table}

\subsection{The Five Steps}

Let $f_{y}(v)=1$ if $v\in F_{y}$ and $0$ otherwise, normalized to zero mean and unit norm over $\mathbb{F}_{q}$; all $F_{y}$ have the same size $r$, so each normalized $f_{y}$ takes two values. DQI with degree $\ell$ and weights $w_{0},\dots,w_{\ell}$ proceeds as follows.

\begin{enumerate}
  \item \emph{Pick a few metrics.} Prepare $\sum_{k\le\ell} w_{k}\binom{m}{k}^{-1/2}\sum_{|e|=k}\prod_{y\in e}\hat f_{y}(e_{y})\lvert e\rangle$, a superposition over patterns $e$ that are nonzero on $k\le\ell$ metrics, where $\hat f_{y}$ is the Fourier transform of $f_{y}$. This is the only place the allowed sets enter.
  \item \emph{Write the syndrome.} Compute $s = B^{\mathsf{T}}e$ into the path register.
  \item \emph{Decode and erase.} A decoder recomputes $e$ from $s$ and subtracts it, which returns the metric register to zero. For OPI the dual code has distance $n+1$, and a Reed--Solomon decoder succeeds for $\ell\le\lfloor n/2\rfloor$. Without this step a Fourier transform of the path register would return a uniformly random path.
  \item \emph{Fourier transform.} Transform each digit of the path register over $\mathbb{F}_{q}$, with $F[x,s]=e^{-2\pi i\,sx/p}/\sqrt p$ for prime $q=p$. The register now holds every path $x$ with amplitude proportional to $\sum_{k}w_{k}\binom{m}{k}^{-1/2}e_{k}\bigl(f_{1}((Bx)_{1}),\dots,f_{m}((Bx)_{m})\bigr)$, with $e_{k}$ the elementary symmetric polynomial; because each $f_{y}$ takes two values, this is a degree-$\ell$ polynomial of the number of metrics $x$ satisfies.
  \item \emph{Measure.} With the weights chosen well, paths that satisfy many metrics are far more likely than under uniform sampling.
\end{enumerate}

The sign convention matters: with it, the amplitude of $x$ is proportional to a polynomial in the values $f_{y}((Bx)_{y})$, and the opposite sign gives a different distribution. The weights $w_{k}$ that maximize the expected score are the top eigenvector of an $(\ell+1)\times(\ell+1)$ matrix. When $2\ell+1<d^{\perp}$, with $d^{\perp}$ the minimum distance of $\ker(B^{\mathsf T})$, that matrix is tridiagonal and instance-independent, and the expected score has the closed form of the finite-size theorem of~\cite{jordan2025}.

\begin{figure}[t]
  \centering
  \includegraphics[width=\columnwidth]{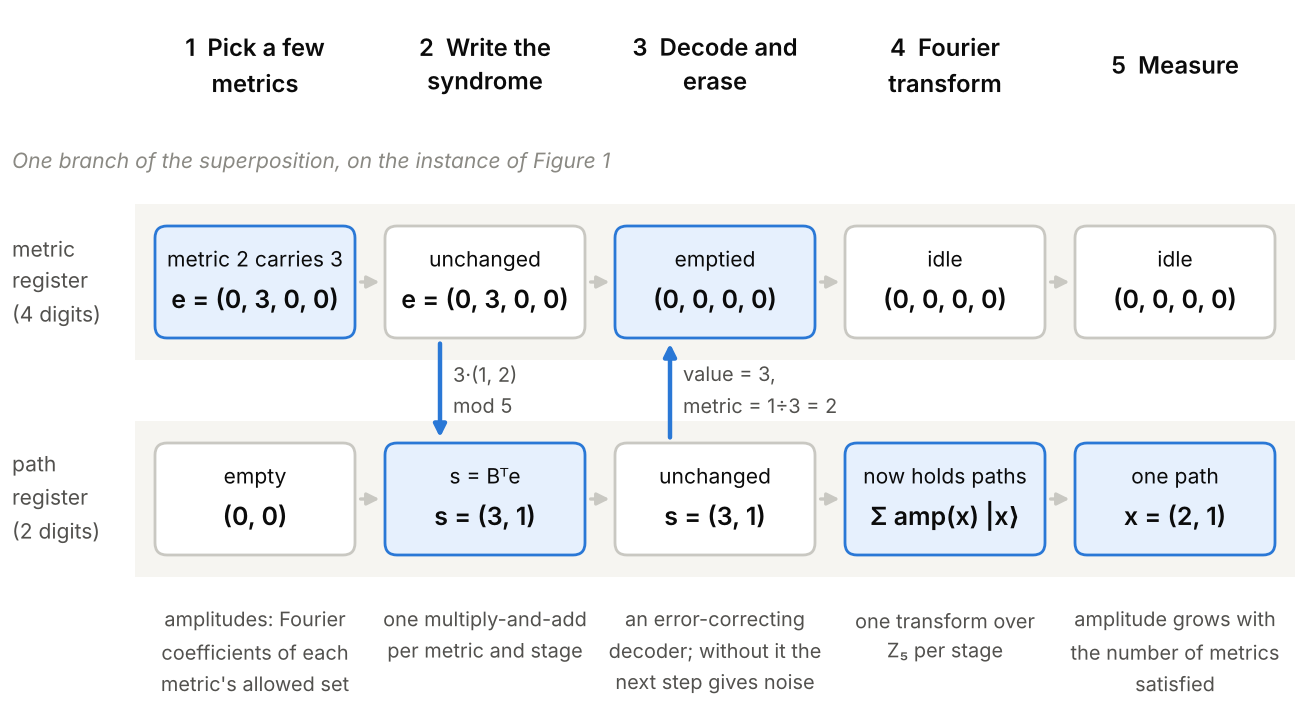}
  \caption{The five steps of DQI on the instance of Fig.~\ref{fig:fabric}, following one branch of the superposition. Blue boxes mark the register each step acts on.}
  \label{fig:steps}
\end{figure}

\begin{figure*}[t]
  \centering
  \includegraphics[width=0.82\textwidth]{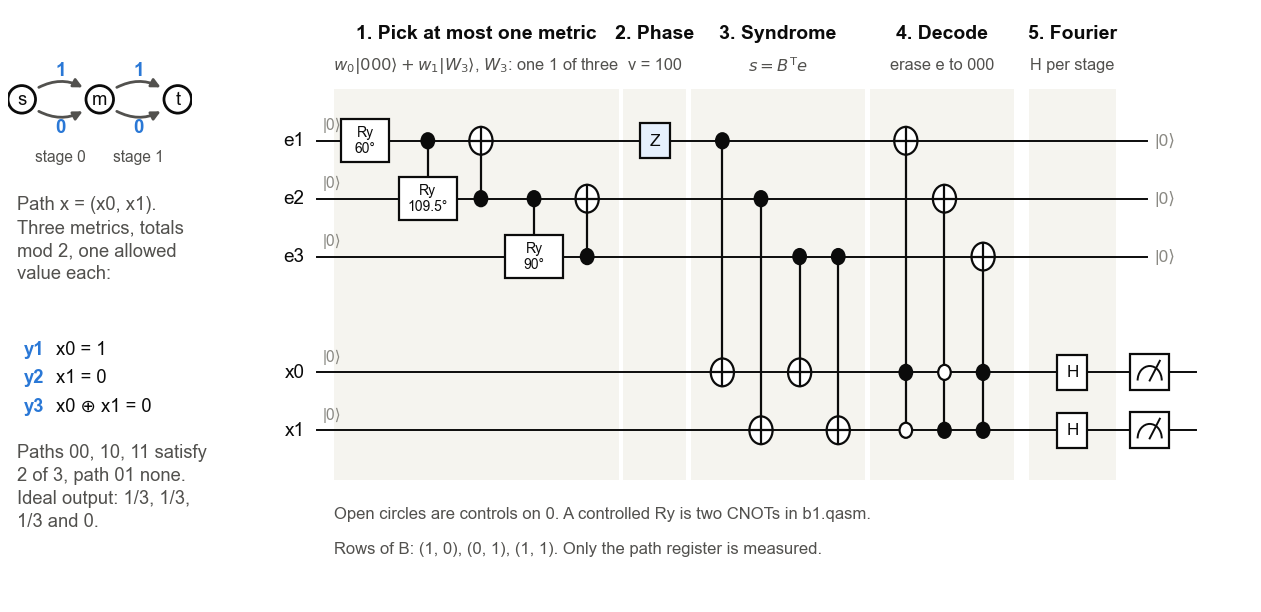}
  \caption{DQI with $\ell = 1$ on the binary instance B1 (two stages, two ports per stage, three metrics with rows $(1,0)$, $(0,1)$, $(1,1)$ of $B$). Metric qubits $e_{1},e_{2},e_{3}$ hold the pattern $e$; path qubits $x_{0},x_{1}$ first receive the syndrome $s = B^{\mathsf{T}}e$ and after the Hadamard gates hold the path. Step 1 prepares $w_{0}\lvert 000\rangle + w_{1}\lvert W_{3}\rangle$, where $\lvert W_{3}\rangle$ is the equal superposition of the three patterns with one picked metric. Step 2 applies the sign of the allowed value $v_{1}=1$. The three Toffoli gates decode the syndrome and erase $e$. Open circles are controls on 0. The OpenQASM file has 10 CNOT-equivalents, counting a controlled rotation as two CNOTs, and 3 Toffoli gates.}
  \label{fig:dqicircuit}
\end{figure*}

The two parts of this paper use the path register in opposite directions. Grover search prepares all paths and then amplifies the good ones at a rate limited by the square root. DQI never prepares the uniform superposition over paths: it builds its state on the metric side, and the linear structure lets a decoder and a Fourier transform move that state onto good paths. The encodings also differ: DQI stores each stage's port as about $\log_{2}q$ qubits, not one qubit per arc. With $q=2$ each stage is one qubit; Fig.~\ref{fig:dqicircuit} draws the smallest hardware instance gate by gate.

\section{Results}

\subsection{Grover Search: Sanity Checks and Baselines}
\label{sec:groverresults}

All results in this subsection are ideal. Five DAGs, I1 to I5, with cost and delay weights from our earlier instance suite, were given a third metric with seeded random weights between 1 and 4. The ranges are the totals of one reference path plus or minus 1, so every instance has a solution and $M$ is known.

Two simulators were used; both only check that the construction is correct. The register-level simulator holds the arc qubits and the three total registers, applies every gate of $U_{S}$ one by one, and applies additions and range tests as arithmetic on registers, not as gates. Its registers are sized by the sum of all arc weights so that no bit string can overflow, which is why I2 needs $2^{25}$ amplitudes although its circuit has 20 qubits. The path-subspace computation applies the ideal oracle and reflection to the $N$ path amplitudes; it agrees with (\ref{eq:sinelaw}) by construction.

\begin{table*}[t]
\caption{Grover search on five instances with three range constraints. Qubits count arcs, total registers and flag. ``Grover 90\,\%'' is the expected number of evaluations to reach 90\,\% success when an attempt with $k$ rounds costs $k$ oracle calls plus one classical check, minimized over $k$; ``Sampling 90\,\%'' counts uniform path samples; ``DP labels'' is the number of labels exact dynamic programming creates.}
\label{tab:groverinstances}
\centering
\footnotesize
\setlength{\tabcolsep}{3.5pt}
\begin{tabular}{lcccccccccccc}
\hline \hline
\textbf{Inst.} & \textbf{Nodes} & \textbf{Arcs} & \textbf{$N$} & \textbf{Ranges} & \textbf{$M$} & \textbf{Qubits} & \textbf{$k^{*}$} & \textbf{$P(0)$} & \textbf{$P(k^{*})$} & \textbf{Grover 90\,\%} & \textbf{Sampling 90\,\%} & \textbf{DP labels} \\
\hline
I1 & 4 & 5 & 3 & 2--4; 9--11; 2--4 & 1 & 18 & 1 & 0.333 & 0.926 & 2 & 5.7 & 4 \\
I2 & 6 & 9 & 5 & 6--8; 5--7; 5--7 & 2 & 20 & 1 & 0.400 & 0.784 & 4 & 4.5 & 8 \\
I3 & 8 & 12 & 8 & 16--18; 20--22; 7--9 & 2 & 28 & 1 & 0.250 & 1.000 & 2 & 8.0 & 15 \\
I4 & 11 & 24 & 27 & 20--22; 17--19; 6--8 & 1 & 39 & 4 & 0.037 & 0.971 & 4 & 61.0 & 38 \\
I5 & 18 & 56 & 256 & 24--26; 25--27; 6--8 & 2 & 74 & 8 & 0.008 & 0.996 & 8 & 293.6 & 96 \\
\hline \hline
\end{tabular}
\end{table*}

\emph{Register level.} I1 and I2 were simulated for three rounds. The probability of a path in range after 0, 1, 2 and 3 rounds is 0.333, 0.926, 0.004, 0.845 on I1 and 0.400, 0.784, 0.077, 0.994 on I2. The deviation from the sine law, the weight left in the total registers after uncomputation and the weight on strings that are not paths are all below $5\times10^{-16}$. This confirms that $U_{S}$, the oracle's arithmetic and its uncomputation fit together; it does not test gate-level adders or comparators.

\emph{Small instances gain nothing.} On I1 to I3 one round is optimal and fewer than ten random samples already succeed with 90\,\% probability; dynamic programming needs at most 15 labels.

\begin{figure*}[t]
  \centering
  \includegraphics[width=0.82\textwidth]{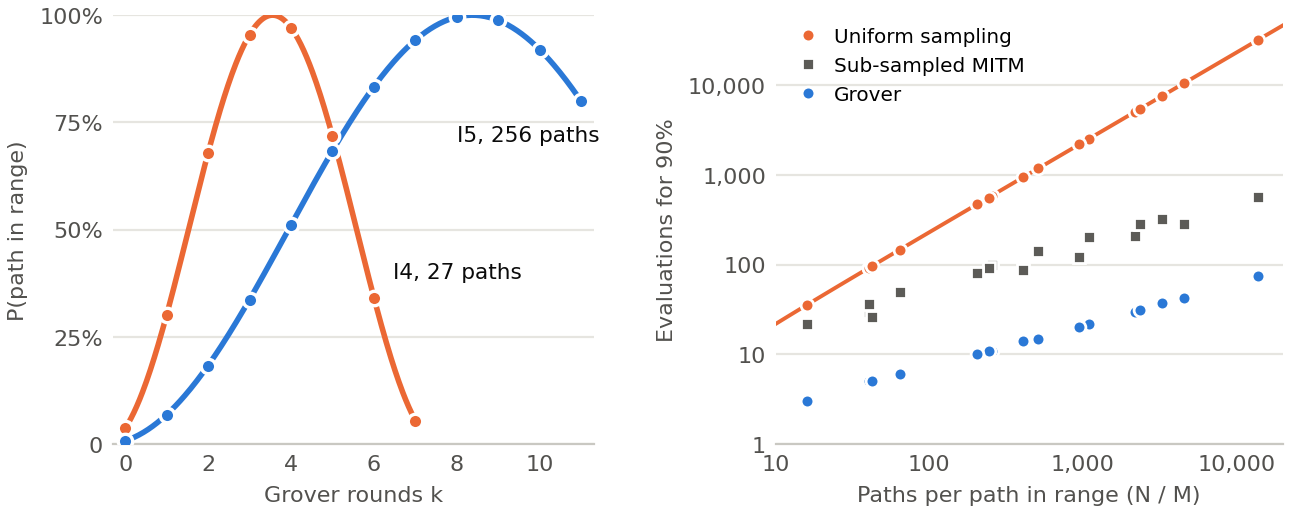}
  \caption{Left: probability that a measurement returns a path in range against the number of Grover rounds, for I4 and I5; points are the path-subspace computation and lines the sine law, which agree by construction. Right: evaluations needed for 90\,\% success on 17 layered DAGs against $N/M$. Uniform sampling and Grover search are computed from closed forms; Grover evaluations count $k$ oracle calls plus one check per attempt. Sub-sampled meet-in-the-middle is measured over 400 trials and counts half-path evaluations; its join tests all pairs, which a range tree does in time nearly linear in the number of halves.}
  \label{fig:groverresults}
\end{figure*}

\emph{Layered DAGs.} Layered DAGs of width $w$ and depth $d$, with $N = w^{d}$ paths and three seeded random metrics, give the scaling series of Table~\ref{tab:layered} and Fig.~\ref{fig:groverresults}. Over the 16 sizes with $k^{*}\ge4$, log-log fits against $N/M$ have slope 0.46 for Grover evaluations and 1.00 for uniform samples. These are properties of the formulas, not measurements. The measured slope of sub-sampled meet-in-the-middle is 0.49: on these product-structured graphs a classical method scales like $\sqrt{N/M}$, within a factor of about eight of Grover search in evaluations. At the largest size, 136 arc qubits and 262{,}144 paths of which 19 are in range, Grover search needs 74 evaluations, sub-sampled meet-in-the-middle 560, exact meet-in-the-middle 1{,}280 half-paths, uniform sampling about 31{,}800, and dynamic programming 18{,}754 labels to decide feasibility exactly.

\begin{table*}[t]
\caption{Grover search and classical baselines on layered DAGs, eight of the 17 sizes. ``MITM sub.\ 90\,\%'' is the number of half-paths sub-sampled meet-in-the-middle needs for 90\,\% success (400 trials); ``MITM exact'' is the number of half-paths exact meet-in-the-middle stores.}
\label{tab:layered}
\centering
\footnotesize
\setlength{\tabcolsep}{3.5pt}
\begin{tabular}{cccccccccccc}
\hline \hline
\textbf{$w$} & \textbf{$d$} & \textbf{Arcs} & \textbf{$N$} & \textbf{$M$} & \textbf{$N/M$} & \textbf{$k^{*}$} & \textbf{Grover 90\,\%} & \textbf{MITM sub.\ 90\,\%} & \textbf{MITM exact} & \textbf{Sampling 90\,\%} & \textbf{DP labels} \\
\hline
2 & 4 & 16 & 16 & 1 & 16 & 3 & 3 & 22 & 8 & 36 & 31 \\
2 & 8 & 32 & 256 & 1 & 256 & 12 & 11 & 100 & 32 & 588 & 253 \\
3 & 6 & 51 & 729 & 3 & 243 & 12 & 11 & 92 & 54 & 558 & 414 \\
4 & 6 & 88 & 4{,}096 & 10 & 410 & 15 & 14 & 86 & 128 & 942 & 1{,}707 \\
3 & 8 & 69 & 6{,}561 & 3 & 2{,}187 & 36 & 30 & 208 & 162 & 5{,}035 & 2{,}586 \\
4 & 8 & 120 & 65{,}536 & 69 & 950 & 24 & 20 & 122 & 512 & 2{,}186 & 10{,}965 \\
3 & 10 & 87 & 59{,}049 & 13 & 4{,}542 & 52 & 43 & 280 & 486 & 10{,}458 & 10{,}340 \\
4 & 9 & 136 & 262{,}144 & 19 & 13{,}797 & 92 & 74 & 560 & 1{,}280 & 31{,}768 & 18{,}754 \\
\hline \hline
\end{tabular}
\end{table*}

\begin{table}[t]
\caption{Logical resources of the Grover circuit under the cost model of Section~\ref{sec:oracle}, without error correction. Toffoli and rotation counts per search use $k^{*}$ rounds. ``Succ.'' is the number of qubits of a successor-choice encoding, $d\lceil\log_{2}\Delta\rceil$; ``DP'' is the number of labels exact dynamic programming creates on the same instance.}
\label{tab:resources}
\centering
\footnotesize
\setlength{\tabcolsep}{2.2pt}
\begin{tabular}{lrrrrrr}
\hline \hline
\textbf{Inst.} & \textbf{Qubits} & \textbf{Succ.} & \textbf{Toff./call} & \textbf{Toff./search} & \textbf{Rot./search} & \textbf{DP} \\
\hline
I1 & 35 & 3 & 119 & 123 & 8 & 4 \\
I2 & 40 & 6 & 151 & 159 & 16 & 8 \\
I3 & 55 & 4 & 323 & 334 & 20 & 15 \\
I4 & 76 & 8 & 561 & 2{,}336 & 224 & 38 \\
I5 & 145 & 10 & 1{,}607 & 13{,}296 & 1{,}248 & 96 \\
$4\times9$ & 310 & 20 & 4{,}669 & 441{,}968 & 36{,}432 & 18{,}754 \\
\hline \hline
\end{tabular}
\end{table}

\subsection{The Grover-Mixer Variant and a Classical Baseline}

Replacing the oracle with one phase rotation per arc qubit and giving the diffusion a free angle turns the circuit into Grover-mixer QAOA (GM-QAOA)~\cite{baertschi2020}. We simulated it on I1 to I5 for the task of minimizing cost under a delay bound. Table~\ref{tab:phaseonly} gives the probability that a measurement returns the constrained optimum. The angles were tuned with knowledge of the optimum, so the numbers show what the circuit can express and are not a search result.

\begin{table}[t]
\caption{GM-QAOA: probability of the constrained optimum, ideal simulation, against a classical Boltzmann sampler at the LARAC multiplier. ``CNOTs'' counts $U_{S}$ only, not a full round.}
\label{tab:phaseonly}
\centering
\footnotesize
\setlength{\tabcolsep}{2.2pt}
\begin{tabular}{lccccccc}
\hline \hline
\textbf{Inst.} & \textbf{Qubits} & \textbf{Paths} & \textbf{CNOTs} & \textbf{Prep.} & \textbf{1 round} & \textbf{2 rounds} & \textbf{Boltzmann} \\
\hline
I1 & 5 & 3 & 9 & 0.333 & 1.000 & 1.000 & 0.500 \\
I2 & 9 & 5 & 16 & 0.200 & 0.743 & 0.980 & 0.246 \\
I3 & 12 & 8 & 25 & 0.125 & 0.505 & 0.919 & 0.468 \\
I4 & 24 & 27 & 63 & 0.037 & 0.147 & 0.319 & 0.493 \\
I5 & 56 & 256 & 169 & 0.004 & 0.015 & 0.030 & 0.114 \\
\hline \hline
\end{tabular}
\end{table}

GM-QAOA at fixed depth is efficiently simulable classically here. A Grover-mixer round changes any amplitude by at most $2\lvert\langle S\vert\psi\rangle\rvert/\sqrt N\le 2/\sqrt N$ and phase separators do not change magnitudes, so after $q$ rounds every path has probability at most $(2q+1)^{2}/N$. The amplitude of any given path, and the normalization, can be computed classically by dynamic programming over the DAG, so rejection sampling from the uniform path distribution accepts with probability at least $(2q+1)^{-2}$. Our classical formula agreed with the subspace simulation to $10^{-15}$ at depths 1, 2, 3 and 5, and the sampler drew 20{,}000 paths for I3 at depth 2 with 7.2 proposals per sample and a total variation distance of 0.0035 from the exact distribution. Under a Grover mixer, paths with equal values of the phased metrics keep equal amplitudes, so the dynamics live in a space whose dimension is the number of distinct value vectors, which is pseudo-polynomial. A classical Boltzmann sampler at the LARAC multiplier finds the optimum with probability 0.49 on I4 and 0.11 on I5, more than two rounds of GM-QAOA.

\subsection{DQI in Simulation}

A script builds the state of each of the five steps as a vector. Decoding is a lookup table, which is valid because the syndromes of all patterns with at most $\ell$ nonzero entries are distinct; the code checks this. Allowed sets are random subsets of $r$ residues. The state is built without knowledge of the optimal path, but the weights $w_{k}$ were computed from a matrix built from the exact score distribution, which is possible only at small size. On the five instances with $2\ell+1<d^{\perp}=n+1$ that matrix equals the instance-independent tridiagonal one, and the DQI mean matches the closed form of~\cite{jordan2025} to $10^{-12}$. On the instance with $n=2$ and $\ell=1$ the condition fails, the dual code has distance 3, and the closed form does not apply.

\begin{table*}[t]
\caption{DQI on six OPI fabrics, state-level simulation. Columns give the mean fraction of metrics satisfied by a random path, by one classical interpolation through $n$ allowed values, by a DQI sample, and by the best path.}
\label{tab:dqi}
\centering
\footnotesize
\begin{tabular}{ccccccccccc}
\hline \hline
\textbf{Ports $p$} & \textbf{Stages $n$} & \textbf{Metrics} & \textbf{Paths} & \textbf{Allowed $r$} & \textbf{$\ell$} & \textbf{$2\ell+1<n+1$} & \textbf{Random} & \textbf{Interpolation} & \textbf{DQI} & \textbf{Best path} \\
\hline
5 & 2 & 4 & 25 & 2 & 1 & no & 0.400 & 0.700 & 0.658 & 0.750 \\
7 & 3 & 6 & 343 & 3 & 1 & yes & 0.429 & 0.714 & 0.643 & 1.000 \\
13 & 5 & 12 & 371{,}293 & 6 & 2 & yes & 0.462 & 0.686 & 0.711 & 1.000 \\
17 & 5 & 16 & 1{,}419{,}857 & 8 & 2 & yes & 0.471 & 0.636 & 0.686 & 1.000 \\
31 & 3 & 30 & 29{,}791 & 15 & 1 & yes & 0.484 & 0.535 & 0.576 & 0.900 \\
61 & 3 & 60 & 226{,}981 & 30 & 1 & yes & 0.492 & 0.517 & 0.556 & 0.767 \\
\hline \hline
\end{tabular}
\end{table*}

On the instance of Fig.~\ref{fig:fabric}, with $\ell = 1$, each of the six best paths has probability 0.119, against 0.040 under uniform sampling; together they take 71.4\,\% of the samples instead of 24.0\,\%.

The per-sample mean of DQI exceeds that of a single interpolation on the four instances with $p\ge13$ and falls below it for $p=5$ and $p=7$. At these sizes that comparison says little. Fluctuations are of order $\sqrt m$, so the best of many interpolations easily beats the DQI mean: for $p=61$ one interpolation satisfies 31.0 metrics on average with standard deviation about 3.8, and the best of 1{,}000 is near 43 under a normal approximation, against a DQI mean of 33.4 and a best path of 46. Several instances also have many perfect paths (for $p=13$ about 35 are expected and 36 exist), where repeated interpolation finds one quickly. Brute force beats both everywhere. The meaningful comparison is the limit $m\to\infty$, where the mean concentrates.

\begin{figure}[t]
  \centering
  \includegraphics[width=\columnwidth]{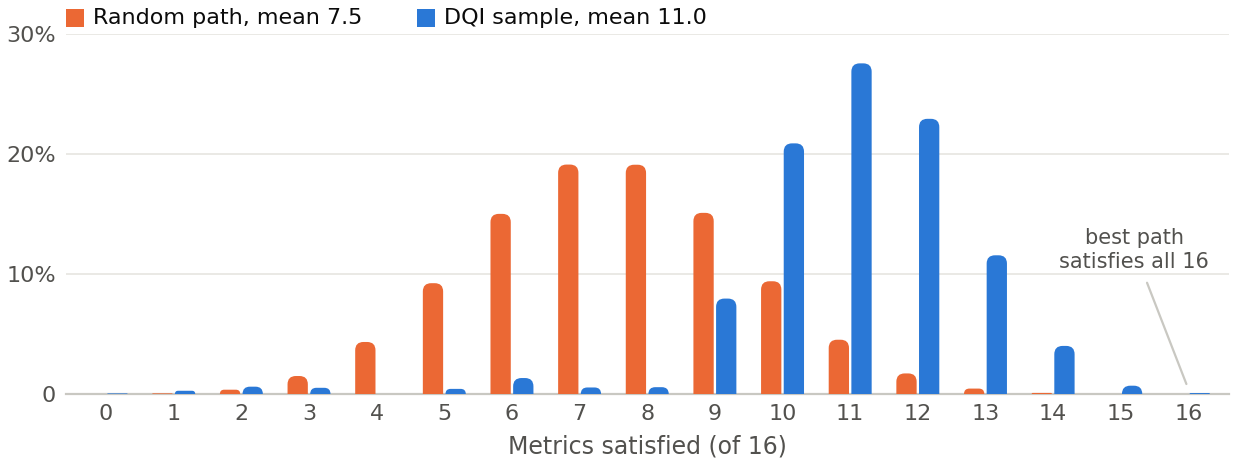}
  \caption{Share of samples by number of metrics satisfied on the fabric with 17 ports and 5 stages (16 metrics, 1.4 million paths). Orange: a uniformly random path. Blue: a DQI sample with $\ell = 2$.}
  \label{fig:scores}
\end{figure}

The closed form can be evaluated at sizes that cannot be simulated (Table~\ref{tab:closedform}); half of the totals are allowed in every row, and every row satisfies $2\ell+1<n+1$.

\begin{table}[t]
\caption{Expected fraction of metrics satisfied at larger sizes, from the finite-size closed form of~\cite{jordan2025}}
\label{tab:closedform}
\centering
\footnotesize
\begin{tabular}{ccccccc}
\hline \hline
\textbf{$p$} & \textbf{$n$} & \textbf{$\ell$} & \textbf{Random} & \textbf{Interp.} & \textbf{DQI} & \textbf{Limit} \\
\hline
101 & 11 & 5 & 0.495 & 0.551 & 0.659 & 0.713 \\
521 & 53 & 26 & 0.499 & 0.550 & 0.695 & 0.717 \\
4{,}099 & 411 & 205 & 0.500 & 0.550 & 0.712 & 0.718 \\
521 & 261 & 130 & 0.499 & 0.750 & 0.919 & 0.933 \\
\hline \hline
\end{tabular}
\end{table}

With about one stage per ten ports, DQI satisfies 72\,\% of the metrics in the limit against 55\,\% for interpolation, the figures Jordan et al.\ put forward as a challenge to classical algorithms~\cite{jordan2025}. With half as many stages as ports the limit is 0.933 against 0.75; there the interpolation baseline $n/m + (1-n/m)/2$ is trivial, and we do not know whether stronger classical algorithms exist at rate $n/m\approx 1/2$. Classical hardness of OPI at these parameters is open; the advantage is evidence, not a theorem.

\subsection{DQI on Hardware}
\label{sec:hw}

The instances of Table~\ref{tab:dqi} need arithmetic modulo $p$ and are too deep for current devices. Two binary fabrics were written as gate-level circuits and run on IBM processors. 
Both are max-XORSAT instances with one metric per nonempty subset of stages, so the rows of $B$ are all nonzero vectors of $\mathbb{F}_{2}^{n}$. They are not OPI.

\emph{B1, five qubits.} Two stages and three metrics; $\ker(B^{\mathsf T})$ is the $[3,1,3]$ repetition code. Three of the four paths satisfy two metrics; the fourth satisfies none. The circuit has 10 CNOT and 3 Toffoli gates (Fig.~\ref{fig:dqicircuit}).

\emph{B2, eleven qubits.} Three stages and seven metrics; $\ker(B^{\mathsf T})$ is the $[7,4,3]$ Hamming code. The path that takes port 1 at every stage satisfies six of the seven metrics. The circuit has 36 CNOT and 9 Toffoli gates and one ancilla qubit, which combines the first two syndrome bits; the decoder is therefore not symmetric in the stages.

\emph{Weights.} Both instances use $\ell=1$ and have $d^{\perp}=3=2\ell+1$, so the hypothesis of the closed form fails and the optimal weights depend on the instance. They need no enumeration. For $\ell=1$ write $S(x)=\sum_{y}f_{y}((Bx)_{y})$ with $f_{y}=\pm1$; then $\mathbb{E}[S]=0$, $\mathbb{E}[S^{2}]=m$ and $\mathbb{E}[S^{3}]=6\sigma$, where $\sigma$ is the number of weight-3 codewords $\{y,y',y''\}$ of $\ker(B^{\mathsf T})$ counted with sign $(-1)^{v_{y}+v_{y'}+v_{y''}}$ for allowed values $v$. The expected value of $S$ is the quadratic form of
\begin{equation}
  A = \begin{pmatrix} 0 & \sqrt{m} \\ \sqrt{m} & 6\sigma/m \end{pmatrix}
\end{equation}
in $(w_{0},w_{1})$, the circuits use its top eigenvector, and the expected number of satisfied metrics is $(m+\lambda_{\max})/2$. B1 has $\sigma=-1$, giving $(w_{0},w_{1})=(0.866,0.5)$ and a mean of 2.00; B2 has $\sigma=1$, giving $(0.648,0.762)$ and 5.05. With the instance-independent weights $(1/\sqrt2,1/\sqrt2)$ the ideal output is different: on B1 each good path gets 31.1\,\% and the forbidden one 6.7\,\%, mean 1.87; on B2 the best path gets 52.2\,\%, mean 5.04. The zero of the forbidden B1 path in Table~\ref{tab:shots} is a property of the tuned weights.

\emph{Checks and runs.} Each circuit is defined once as a gate list that feeds a numpy simulator and an OpenQASM writer. All 35 checks of the gate lists passed before the runs, and a second script that shares no code with the first, with its own QASM reader and its own reference from the definition of the DQI state, reproduced the ideal distributions to $2\times10^{-16}$ and confirmed that all unmeasured qubits return to zero. Each OpenQASM file was transpiled with Qiskit and run once with 4{,}096 shots. Only the path qubits were measured. Table~\ref{tab:hardware} lists what was recorded; the transpiler optimization level and seed, the Sampler options and the job identifiers were not retained.

\begin{table}[t]
\caption{Hardware runs. Error figures are medians over the whole device on the day of the runs (Appendix~\ref{app:calibration}), not values of the qubits used.}
\label{tab:hardware}
\centering
\footnotesize
\setlength{\tabcolsep}{3pt}
\begin{tabular}{p{0.3\columnwidth}p{0.29\columnwidth}p{0.29\columnwidth}}
\hline \hline
\textbf{Property} & \textbf{B1} & \textbf{B2} \\
\hline
Backend & ibm\_marrakesh (Heron r2) & ibm\_fez (Heron r2) \\
Shots, runs & 4{,}096, one & 4{,}096, one \\
Qubits in the circuit & 5 & 11 \\
Gates as written & 10 CNOT, 3 Toffoli & 36 CNOT, 9 Toffoli \\
Gates after transpiling & 44 CZ, 89 SX, 70 RZ, 3 X & 179 CZ, 370 SX, 222 RZ, 5 X \\
Physical qubits (from the transpiler drawing) & 137, 146--149 & 68, 69, 77, 78, 85--90, 97 \\
Median CZ error & 0.24\,\% & 0.24\,\% \\
Two-qubit error, layered & 0.50\,\% & 0.43\,\% \\
Median readout error & 1.66\,\% & 1.22\,\% \\
Median $T_{1}$, $T_{2}$ & 164, 93 $\mu$s & 134, 100 $\mu$s \\
\hline \hline
\end{tabular}
\end{table}

Each shot returns one path. Table~\ref{tab:shots} gives the share of shots per path next to two references computed before the runs: the ideal circuit with tuned weights, and a noise model with 1\,\% depolarizing error after every two-qubit gate of the circuit as written (28 and 90 CNOT-equivalents, six per Toffoli) and 2\,\% readout error. That model ignores the routing overhead of transpilation, idling and decoherence.

\begin{table}[t]
\caption{Share of shots per path. With 4{,}096 shots the standard error of a share is at most 0.8 percentage points.}
\label{tab:shots}
\centering
\footnotesize
\begin{tabular}{lccccc}
\hline \hline
\textbf{Inst.} & \textbf{Path} & \textbf{Metrics} & \textbf{Ideal} & \textbf{Noise model} & \textbf{Hardware} \\
\hline
B1 & 0, 0 & 2 of 3 & 33.33\,\% & 30.9\,\% & 29.1\,\% \\
B1 & 1, 0 & 2 of 3 & 33.33\,\% & 31.6\,\% & 35.8\,\% \\
B1 & 1, 1 & 2 of 3 & 33.33\,\% & 30.6\,\% & 26.8\,\% \\
B1 & 0, 1 & 0 of 3 & 0.00\,\% & 6.9\,\% & 8.3\,\% \\
B2 & 1, 1, 1 & 6 of 7 & 54.46\,\% & 31.2\,\% & 19.1\,\% \\
B2 & 0, 1, 0 & 4 of 7 & 10.95\,\% & 11.1\,\% & 8.1\,\% \\
B2 & 0, 1, 1 & 4 of 7 & 10.95\,\% & 12.7\,\% & 18.9\,\% \\
B2 & 1, 0, 0 & 4 of 7 & 10.95\,\% & 10.9\,\% & 9.9\,\% \\
B2 & 1, 0, 1 & 4 of 7 & 10.95\,\% & 12.4\,\% & 11.3\,\% \\
B2 & 0, 0, 0 & 2 of 7 & 0.58\,\% & 5.9\,\% & 10.2\,\% \\
B2 & 0, 0, 1 & 2 of 7 & 0.58\,\% & 6.9\,\% & 12.5\,\% \\
B2 & 1, 1, 0 & 2 of 7 & 0.58\,\% & 8.9\,\% & 10.0\,\% \\
\hline \hline
\end{tabular}
\end{table}

\begin{figure}[t]
  \centering
  \includegraphics[width=\columnwidth]{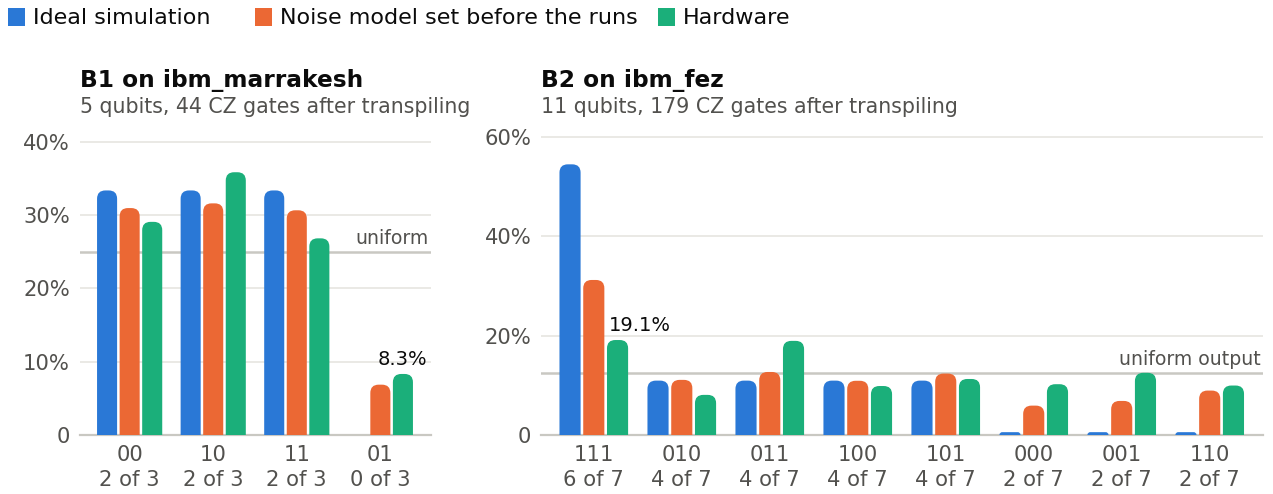}
  \caption{Share of shots per path for B1 and B2: ideal circuit with tuned weights, the noise model set before the runs, and hardware. Under each path is the number of metrics it satisfies.}
  \label{fig:hardware}
\end{figure}

\begin{table}[t]
\caption{Summary of each run. Distance is total variation distance. Gain is the mean number of metrics satisfied minus its value for uniform output; ``gain kept'' is the hardware gain over the ideal gain.}
\label{tab:summary}
\centering
\footnotesize
\begin{tabular}{lcccc}
\hline \hline
\textbf{Quantity} & \textbf{Uniform} & \textbf{Ideal} & \textbf{Noise} & \textbf{Hardware} \\
\hline
B1: mean metrics, of 3 & 1.50 & 2.00 & 1.86 & 1.83 \\
B1: forbidden path share & 25.0\,\% & 0.0\,\% & 6.9\,\% & 8.3\,\% \\
B1: distance from ideal & 0.250 & 0 & 0.069 & 0.108 \\
B1: ideal gain kept & 0\,\% & 100\,\% & 73\,\% & 67\,\% \\
B2: mean metrics, of 7 & 3.50 & 5.05 & 4.19 & 3.73 \\
B2: best path share & 12.5\,\% & 54.5\,\% & 31.2\,\% & 19.1\,\% \\
B2: distance from ideal & 0.420 & 0 & 0.233 & 0.393 \\
B2: ideal gain kept & 0\,\% & 100\,\% & 44\,\% & 15\,\% \\
\hline \hline
\end{tabular}
\end{table}

\emph{The pattern survives on five qubits.} On B1 the forbidden path receives 8.3\,\% of the shots instead of the 25\,\% of uniform output. A shot satisfies 1.83 metrics on average, against 1.50 for uniform output and 2.00 for the ideal circuit. The difference from uniform is 24.6 standard errors of the uniform null distribution (38.6 with the empirical standard error). Against the instance-independent weights the device would have kept 91\,\% of the gain, but those are not the weights it ran.

\emph{Most of it is lost on eleven qubits.} On B2 the best path receives 19.1\,\% of the shots, against 54.5\,\% for the ideal circuit and 12.5\,\% for uniform output. Its excess over uniform is 12.8 binomial null standard errors, so a signal is present, but the output is closer to uniform (distance 0.13) than to ideal (0.39). The kept gain is 15\,\% against either weight choice.

\emph{One effective error rate fits both runs.} Suppose every transpiled CZ gate fails with probability $p_{\mathrm{eff}}$ and a failure randomizes the output; then the kept gain is $(1-p_{\mathrm{eff}})^{n_{\mathrm{CZ}}}$. Matching the runs gives $p_{\mathrm{eff}}=0.92\,\%$ for B1 (44 CZ) and $1.07\,\%$ for B2 (179 CZ); with $p_{\mathrm{eff}}=1\,\%$ the model predicts 64\,\% and 17\,\%, against 67\,\% and 15\,\% measured. The difference between B1 and B2 is thus consistent with circuit size alone, without assuming a faulty qubit; two runs and one fitted parameter cannot establish more. The effective rate is about four times the median CZ error; the device medians predict error-free shots in 82\,\% and 55\,\% of cases, or 73\,\% and 39\,\% with the layered two-qubit error. Idling during routing, crosstalk, readout error and the error of the qubits actually used can each contribute to the excess, and these runs cannot separate them.

\emph{The loss is uneven across stages.} In the ideal output of B2 each stage takes port 1 with probability 76.9\,\%. The device gives 50.2\,\% at stage 0, 56.1\,\% at stage 1 and 61.9\,\% at stage 2. Several explanations fit. The logical circuit is not stage-symmetric: the stage-0 qubit takes part in eight two-qubit gates and two Toffoli gates, mostly as a control of the ancilla, and stage 2 in four and seven. Routing SWAPs may act asymmetrically, the stage-0 qubit may idle longer before its final Hadamard, and an incompletely erased metric register would leave the path qubits entangled. Measuring all qubits, which should return the metric register and the ancilla to zero, would separate failed decoding from decoherence of the path register and allow post-selection.

Both instances can be solved by inspection; these runs show that the interference pattern of a small DQI circuit survives on a current device in part and say nothing about advantage.

\section{Conclusion and Limitations}
\label{sec:limits}

Multi-constrained path finding can be run as a Grover search on any DAG with a provably correct state preparation that never leaves the set of paths, and in a special algebraic case it can be handed to DQI. The first gives $O(\sqrt{N/M})$ oracle calls, an exponent that classical meet-in-the-middle matches in the worst case and that quantum algorithms with quantum memory beat, at a gate cost far above dynamic programming on every instance we tested. The second inherits an advantage that is supported by evidence, not proven, for a problem that is far from ordinary routing. The limitations are the following.

\subsection{General Case and Grover Search}
\begin{enumerate}
  \item \emph{No practical advantage.} Plain Grover search is not the best known quantum algorithm for this problem, and the instances tested are pseudo-polynomial, where dynamic programming wins by orders of magnitude.
  \item \emph{Simulation only, and sanity checks only.} All Grover results are ideal. The register-level simulation applies arithmetic as register operations, and the path-subspace computation agrees with the sine law by construction. No noise was modeled and nothing was run on hardware.
  \item \emph{Resource estimate, not synthesis.} Table~\ref{tab:resources} counts logical gates under a stated cost model; the oracle was not compiled to gates and simulated, and error-correction overhead is not included.
  \item \emph{Easy instances.} The ranges were placed around a known path, so $M\ge1$ is known. Weights are small. Instances with large weights, unknown $M$, infeasibility, sparse irregular DAGs and real topologies after hop layering are not tested, and search with unknown $M$ and the minimization variant are described but not simulated.
  \item \emph{Acyclic graphs only.} General graphs need a hop-bounded layering. With non-negative weights and upper bounds only, walks are dominated by simple paths and this is harmless; with lower bounds a walk can satisfy a range that no simple path satisfies, so the reduction is incorrect, and enforcing simplicity in superposition is expensive.
  \item \emph{Classical comparison partly empirical.} The sub-sampled meet-in-the-middle bound is measured on product-structured DAGs, where joining any two halves gives a path; on general DAGs halves must meet at a common node.
\end{enumerate}

\subsection{Special Case and DQI}
\begin{enumerate}
  \item \emph{Not a new algorithm.} The OPI dictionary is a renaming, and the graph plays no role once weights must be node-independent (Remark~\ref{rem:linear}).
  \item \emph{Artificial weights.} Delay and cost are not taken modulo a prime, and we have not identified a networking application with power-of-index metrics; periodic scheduling is a candidate for linear modular metrics in general.
  \item \emph{Different objective.} DQI maximizes the number of satisfied metrics. Satisfying all of a set of ordinary integer budgets on the same fabric contains subset sum.
  \item \emph{Advantage not proven.} Classical hardness of OPI at the parameters of interest is open. The proven separation is for a folded variant in an oracle model~\cite{kramer2026}, and DQI is not optimal on worst-case instances~\cite{sun2026,horinaga2026}.
  \item \emph{Ranges may be easier.} Contiguous allowed sets form noisy polynomial reconstruction, where lattice methods succeed in some regimes~\cite{bleichenbacher2000}; we have not checked whether they do at the parameters where DQI is expected to win.
  \item \emph{Small simulations.} Every simulated instance can be enumerated, and brute force beats DQI on all of them. The simulated weights were computed from the exact score distribution.
  \item \emph{Hardware instances are binary, trivial and outside OPI.} B1 and B2 are max-XORSAT with $\ell=1$ and $d^{\perp}=2\ell+1$, solved by inspection. The five-level circuit for Fig.~\ref{fig:fabric} was only simulated.
  \item \emph{Single, uncontrolled, partly documented runs.} Each circuit was run once on a different backend, without the control circuits (shifted targets; no decoding step) that would attribute the suppression of bad paths to DQI interference, without error suppression we can confirm, and with only the path register measured. Transpiler settings, Sampler options, job identifiers and per-qubit calibration were not retained, and the B2 layout is read from the transpiler drawing. Run-to-run variation is unknown.
\end{enumerate}

The next steps we consider most useful are the two control circuits and five or more repeated runs on one backend with recorded layout, seeds and per-qubit calibration, with all qubits measured; runs of $U_{S}$ on I1 to I3 with post-selection; gate-level synthesis and simulation of the oracle; tests on instances with large weights; and a test of whether contiguous ranges make the special case easier classically.

\section*{AI Disclosure}
Generative AI tools, such as large language models, were used to assist with writing and running the simulation and analysis code and with drafting and editing the text. The author reviewed and verified the results and takes full responsibility for the content.


\appendices

\section{Code and Data}
\label{app:code}

The simulations, circuits and analyses are in the public repository \url{https://github.com/duman190/q_path_finder}. Every script runs from a checkout with Python 3 and numpy.
\begin{itemize}
  \item \texttt{qpathfinder/}: state preparation, the DAG instances, GM-QAOA, LARAC and the Boltzmann sampler.
  \item \texttt{grover\_mcp/grover\_mcp.py}: Grover search with range constraints, both simulators, Tables~\ref{tab:groverinstances} and~\ref{tab:layered}.
  \item \texttt{grover\_mcp/baselines.py}: dynamic programming, exact and sub-sampled meet-in-the-middle, evaluations to 90\,\%, and the resource estimate of Table~\ref{tab:resources}; \texttt{figures.py} draws Figs.~\ref{fig:circuit} and~\ref{fig:groverresults}.
  \item \texttt{dqi\_paths/}: state-level DQI simulation and the closed form.
  \item \texttt{dqi\_cirq/}: gate lists and OpenQASM files of the DQI circuits, the checks, the independent re-implementation \texttt{independent\_check.py}, and the drawing of Fig.~\ref{fig:dqicircuit}. Cirq builders are included but were not executed.
  \item \texttt{hardware/}: raw counts, \texttt{score\_runs.py} for Tables~\ref{tab:shots} and~\ref{tab:summary}, and \texttt{reanalysis.py} for the weight matrix, both weight choices, the null and empirical standard errors, the effective error per transpiled CZ and the per-stage gate counts. Screenshots of the backend pages and of the transpiled circuits are in \texttt{ieee\_paper/figures/}.
\end{itemize}

\section{Hardware Calibration Records}
\label{app:calibration}

Table~\ref{tab:calib} lists the medians printed on the IBM backend pages on the day of the runs. Both devices are 156-qubit Heron r2 systems. The pages do not list the error of the particular qubits each circuit was mapped to, and per-qubit calibration was not retrieved.

\begin{table}[h]
\caption{Device medians on the day of the runs}
\label{tab:calib}
\centering
\footnotesize
\begin{tabular}{lcc}
\hline \hline
\textbf{Quantity} & \textbf{ibm\_marrakesh (B1)} & \textbf{ibm\_fez (B2)} \\
\hline
Median CZ error & 0.237\,\% & 0.244\,\% \\
Layered two-qubit error & 0.501\,\% & 0.431\,\% \\
Best pair & 0.116\,\% & 0.123\,\% \\
Median SX error & 0.041\,\% & 0.028\,\% \\
Median readout error & 1.66\,\% & 1.22\,\% \\
Median $T_{1}$ & 163.7 $\mu$s & 134.3 $\mu$s \\
Median $T_{2}$ & 92.9 $\mu$s & 99.7 $\mu$s \\
\hline \hline
\end{tabular}
\end{table}

\end{document}